\documentclass[12pt]{iopart}

\usepackage{iopams}  
\usepackage{color}
\usepackage{graphics}
\usepackage{harvard}

\newcommand{\grad}{\nabla}

\newcommand{\wz}{\omega}
\newcommand{\wzp}{\omega_{+}}
\newcommand{\wzm}{\omega_{-}}
\newcommand{\wzpm}{\omega_{\pm}}
\newcommand{\hwz}{\hat{\omega}}
\newcommand{\hwzp}{\hat{\omega}_{+}}
\newcommand{\hwzm}{\hat{\omega}_{-}}
\newcommand{\hwzpm}{\hat{\omega}_{\pm}}
\newcommand{\dwz}{\delta\omega}
\newcommand{\vr}{\vec{r}}
\newcommand{\vk}{\vec{k}}

\newcommand{\vq}{\vec{q}}
\renewcommand{\div}{\nabla \cdot}
\newcommand{\vu}{\vec{u}}

\renewcommand{\vec}[1]{\bi{#1}}

\begin{document}

\title[]{Mechanism of self-organization in point vortex system}

\author{Yuichi Yatsuyanagi$^1$\footnote{Corresponding 
author: yatsuyanagi.yuichi@shizuoka.ac.jp} and Tadatsugu Hatori$^2$}

\address{$^1$Faculty of Education, Shizuoka University, Suruga-ku, Shizuoka 422-8529, JAPAN}
\address{$^2$National Institute for Fusion Science, Toki, Gifu 509-5292, JAPAN}

\begin{abstract}
A mechanism of the self-organization in an unbounded two-dimensional (2D) point 
vortex system is discussed.
A kinetic equation for the system with positive and negative vortices is 
derived using the Klimontovich formalism.
Similar to the Fokker-Planck collision term, the obtained collision 
term consists of a diffusion term and a drift term.
It is revealed that the mechanism for the self-organization in the 2D point vortex
system at negative absolute temperature is mainly provided by the drift term.
Positive and negative vortices are driven toward opposite directions respectively 
by the drift term.
As a result, well-known, two isolated clumps with positive and negative 
vortices, respectively, are formed as an equilibrium distribution.
Regardless of the number of species of the vortices, either single- or double-sign,
it is found that the collision term has following physically good properties:
(i) When the system reaches a quasi-stationary state near the thermal equilibrium state
with negative absolute temperature, the sign of $d \omega / d \psi$ is expected 
to be positive, where $\omega$ is the vorticity and $\psi$ is the stream function.
In this case, the diffusion term decreases the mean field energy, while the drift term
increases it.
As a whole, the total mean field energy is conserved.
(ii) Similarly, the diffusion term increases the Boltzmann entropy, while the
drift term decreases it.
As a whole, the total entropy production rate is positive or zero ($H$ theorem),
which ensures that the system relaxes to the global thermal equilibrium state
characterized by the zero entropy production.
\end{abstract}

\vspace{2pc}
\noindent{\it Keywords}: Point vortex system, Negative absolute temperature, Self-organization

\maketitle

\section{Introduction\label{sec:introduction}}
In this paper, we propose a general mechanism of the self-organization for the two-dimensional (2D) point vortex 
system composed of double-sign vortices
through a newly obtained kinetic equation.
The kinetic equation clearly elucidates the mechanism of the self-organization,
in other words, a condensation of the same-sign vortices and a separation
of the different-sign vortices.

At first, let us briefly introduce a hierarchy of the plasma kinetic equations.
There are several equations with different scales.
The most microscopic equation is the Klimontovich equation which describes 
a time evolution of a microscopic phase density $\hat{f}_{\alpha}(\vr, \vec{v},t)$ 
for the $\alpha$-th plasma species with the charge $q_{\alpha}$ and the mass $m_{\alpha}$
in a six-dimensional phase space
\begin{equation}
	\frac{\partial \hat{f}_{\alpha} (\vr, \vec{v},t)}{\partial t} + \vec{v} \cdot \grad \hat{f}_{\alpha} (\vr, \vec{v},t)
	+ \frac{q_{\alpha}}{m_{\alpha}}(\hat{\vec{E}}(\vr, t) + \vec{v} \times \hat{\vec{B}}(\vr, t))\cdot \frac{\partial \hat{f}_{\alpha} (\vr, \vec{v},t)}{\partial \vec{v}} = 0 \label{eqn:Klimontovich}
\end{equation}
where $\hat{\vec{E}}(\vr, t)$ and $\hat{\vec{B}}(\vr, t)$ are the microscopic electric and magnetic fields.
Equation (\ref{eqn:Klimontovich}) has a formal discretized solution
\begin{equation}
	\hat{f}_{\alpha}(\vr, \vec{v},t) = \sum_i^{N_{\alpha}}\delta(\vr - \vr_{\alpha,i}(t)) \delta(\vec{v} - \vec{v}_{\alpha,i}(t)) \label{eqn:solKlimontovich}
\end{equation}
where the position vector and the velocity of the $i$-th particle of the $\alpha$-th component
are given by $\vr_{\alpha,i}$ and $\vec{v}_{\alpha,i}$, respectively.
Number of particles of the $\alpha$-th component is given by $N_{\alpha}$.

It is difficult to make direct use of the microscopic equation 
(\ref{eqn:Klimontovich}) because of its complexity. 
We shall therefore proceed to the ensemble-average. 
It is assumed that the microscopic phase density $\hat{f}_{\alpha} (\vr, \vec{v}, t)$ 
is composed of a macroscopic phase density $f_{\alpha}(\vr, \vec{v}, t)$ and 
a fluctuation $\delta f_{\alpha}(\vr, \vec{v},t)$
\begin{eqnarray}
	\hat{f}_{\alpha} (\vr, \vec{v}, t) & = & f_{\alpha} (\vr, \vec{v}, t) + \delta f_{\alpha} (\vr, \vec{v}, t), \\
	f_{\alpha} (\vr, \vec{v}, t) & \equiv &  \langle \hat{f}_{\alpha} (\vr, \vec{v}, t) \rangle
\end{eqnarray}
where the operator $\langle \cdot \rangle$ means the ensemble average.
In the same manner, the other physical quantities are rewritten into 
the averaged value and the fluctuation.
Inserting the above expressions into the microscopic equation (\ref{eqn:Klimontovich}) 
and averaging the equation, we obtain the following macroscopic equation 
with a collisional effect in the right hand side:
\begin{equation}
	\frac{\partial f_{\alpha}}{\partial t} + \vec{v} \cdot \grad f_{\alpha} + \frac{q_{\alpha}}{m_{\alpha}}(\vec{E} + \vec{v} \times \vec{B}) \cdot \frac{\partial f_{\alpha}}{\partial \vec{v}} = \frac{q_{\alpha}}{m_{\alpha}} \left\langle\left(\delta \vec{E} + \vec{v} \times \delta\vec{B} \right) \cdot \frac{\partial \delta f_{\alpha}}{\partial \vec{v}}  \right\rangle. \label{eqn:withcollision}
\end{equation}
This equation describes a time evolution of a system in terms 
of the continuous probability density function $f_{\alpha}$ instead of the 
discretized microscopic phase density $\hat{f}_{\alpha}$.
Expressing the collision term in a form of the perturbation expansion and
gathering terms of the appropriate order, the Fokker-Planck type equation 
for a plasma is obtained
\begin{equation}
	\frac{\partial f_{\alpha}}{\partial t} + \vec{v} \cdot \grad f_{\alpha} + \frac{q_{\alpha}}{m_{\alpha}}(\vec{E} + \vec{v} \times \vec{B}) \cdot \frac{\partial f_{\alpha}}{\partial t} = \frac{\partial }{\partial \vec{v}} \cdot \left( {\sf D} \cdot \frac{\partial f_{\alpha}}{\partial \vec{v}} + \vec{A} f_{\alpha} \right) \label{eqn:Fokker-Planck}
\end{equation}
where $\sf D$ is a diffusion tensor and $\vec{A}$ is a friction. 
The above procedure is called the Klimontovich formalism \cite{Klimontovich}. 
In plasmas, long-range Coulomb interactions rather than collisions govern 
a whole dynamics of a system. 
For these systems, the Vlasov equation is appropriate, which is obtained by 
dropping the collision term in (\ref{eqn:Fokker-Planck}):
\begin{equation}
	\frac{\partial f_{\alpha}}{\partial t} + \vec{v} \cdot \grad f_{\alpha} + \frac{q_{\alpha}}{m_{\alpha}}(\vec{E} + \vec{v} \times \vec{B}) \cdot \frac{\partial f_{\alpha}}{\partial t} = 0. \label{eqn:Vlasov}
\end{equation}
Namely, (\ref{eqn:Vlasov}) is a collisionless equation by approximation.

We have noticed that the same hierarchy exist in the 2D Euler equation. 
The point vortex solution is a counterpart of (\ref{eqn:solKlimontovich})
and we assume that this solution is a microscopic one. 
Therefore, we regard the 2D Euler equation which has the point vortex 
solution as a microscopic equation. 
Applying the Klimontovich formalism to the microscopic Euler equation, 
we will obtain a corresponding macroscopic equation to (\ref{eqn:Fokker-Planck}) 
with a collisional effect. 
By dropping the collision term from the obtained equation, 
we will obtain the inviscid 2D Euler equation in the usual sense.

The 2D point vortex system has been
successfully applied to understand the various phenomena
including 2D turbulence \cite{Kida1985,Eyink,Tabeling,Kraichnan}, 
neutral \cite{TaylorMcNamara1971} and nonneutral \cite{DubinJin2001,Yatsuyanagi2003-2} plasmas.
These phenomena share a common keyword, ``self-organization".
In the context of the self-organization, possibility of the negative 
temperature state in the 2D point vortex system was first pointed out by 
\citeasnoun{Onsager}. 
The concept of the negative temperature state is convenient to explain 
how a large scale structure, such as Jupiter's Great Red Spot and typhoons,
is formed before stored energy is exhausted by a dissipative process.
If the temperature is negative, no spatially homogeneous thermal 
equilibrium distribution exists. 
Such states have been discussed in several ways.
\citeasnoun{Joyce} derived the sinh-Poisson equation which 
determines the thermal equilibrium distribution of double-sign 
point vortices. 
\citeasnoun{Kida1975} discussed the axisymmetric equilibrium distribution of a single- 
and double-sign vortices bounded in a circular domain using the 
well-known maximizing entropy techniquie.

To understand the relaxation process toward such thermal equilibrium states, 
it is necessary to develop a kinetic theory. 
A kinetic equation with a collisional effect describes how a system relaxes 
to an equilibrium state.
Kinetic theory of the point vortex system has attracted a lot of attention.
A general kinetic equation for the point vortex system has been obtained 
by Chavanis with several methods including projection operator, the BBGKY hierarchy and
the Klimontovich formalism \cite{Chavanis2001,Chavanis2008}.
The kinetic equations have a Fokker-Planck type collision term
that is composed of a diffusion term and a drift term.
The drift term was first evidenced in \citeasnoun{Robert1992}
and \citeasnoun{Chavanis1998}.
A kinetic theory for multi-species point vortex system was discussed by
\citeasnoun{DubinONeil1988} and \citeasnoun{Dubin2003} in the context of 
magnetized plasmas with the Klimontovich formalism.
\citeasnoun{Chavanis2007} discussed the axisymmetric case with the Fokker-Planck type
collision term.
The result has an issue that a relaxation process stops before
the system reaches a Boltzmann-type thermal equilibrium state, if
a profile of an angular velocity is a monotonic decay function.

In the previous paper \cite{Yatsuyanagi2015}, we have derived a kinetic 
equation having a Fokker-Planck type collision term
for a single-species point vortex system with a weak mean flow.
We have paid a special attention to treat a weak mean flow case correctly, 
which is a complementary case to many works by 
\citeasnoun{Chavanis1998}, \citeasnoun{Chavanis2001}, \citeasnoun{Chavanis2008}, 
\citeasnoun{DubinONeil1988} and \citeasnoun{Dubin2003}.
The phrase ``weak mean flow" means that the number of the point vortices 
$N$ has a lower and an upper limits,
\begin{equation}
	1 < \pi \left(\frac{R}{L}\right)^2 < N < \frac{\pi}{16}\left(\frac{R}{L}\right)^4 \label{eqn:limit}
\end{equation}
where $R$ is a characteristic system size and $L$ is a characteristic microscopic size.
See Appendix in \citeasnoun{Yatsuyanagi2015} for detail and we will present a refined estimation
in section \ref{subsec:epsilon}. 
With this limit, the approximation that a mean trajectory is linear in the 
microscopic time scale is validated. 
The obtained collision term has the Fokker-Planck form, namely, it is composed 
of the diffusion term and the drift term. 
It was revealed that the diffusion term dissipate the mean field energy, 
while the drift term increases it. 
As a whole, the total men field energy is conserved. 
In other words, the drift term accumulates the vortices in the same place, 
while the diffusion term disperses the accumulated vortices. 
In addition, the collision term exhibits several physically important properties: 
(a) it includes a nonlocal effect; 
(b) it satisfies the $H$ theorem; 
(c) its effect vanishes in the thermal equilibrium state. 
This means that in contrast to \citeasnoun{Chavanis2007}, 
the kinetic equation ensures that a system relaxes to a Boltzmann 
thermal equilibrium state even if the profile is a 
monotonically-decaying symmetric one.

The most remarkable feature of the self-organization in a 2D system 
is a large-scale vortex formation with the same-sign vorticity, which 
is expected to be connected with the inverse-cascade in the 2D turbulence. 
In such systems, it is quite common that positive and negative vortices coexist.
However, the above-mentioned single-species model cannot handle a system 
with positive and negative vortices.
Thus, to understand the self-organization process in a system in which vortices 
with the clockwise direction and with the counterclockwise direction coexist, 
we need to extend the previous single-species model to a double-species 
model. 
So, in this paper, we present a new model for a double-species point 
vortex system with a weak mean flow.

In a single-species point vortex system, a remarkable feature of the 
self-organization is a condensation of the same-sign vortices, 
although no attractive forces act between them.
The previous single-species model can describe the condensation 
of the single-sign vortices correctly.
In a double-species point vortex system, an additional 
remarkable feature of the self-organization appears. 
That is a ``charge-separation" of positive and negative vortices. 
Namely, positive vortices isolate themselves from negative vortices 
and are condensed into a clump which is exclusively composed of positive vortices, 
and vice versa. 
We emphasize that a new finding in the double-species model is that 
the model can explain the feature of charge-separation brought 
by the drift term in addition to the clumping feature which is also 
provided by the drift term regardless of the single- or double-species 
system.
Thus, the current model bears discussions for the 2D turbulence as the 
important features of the 2D turbulence, the condensation and the 
charge-separation, are incorporated in it.
We will also demonstrate the obtained collision term has physically 
important properties similar to the single-species model.

The organization of this paper is as follows.
In section 2, the point vortex system and a kinetic equation are briefly introduced.
We demonstrate explicit formulae for the diffusion term and the drift
term.
In section 3, physical properties of the collision term are examined.
The important role of the drift term
in the 2D self-organization with negative absolute temperature will be discussed.
In section 4, we present a numerical result of the self-organization
of the point vortices.
Finally in sections 4 and 5, we give a discussion and a conclusion.
\section{Kinetic equation for 2D point vortex system}
Let us consider a 2D point vortex system consisting of $N_+$ positive and $N_-$ negative vortices \cite{Newton},
\begin{eqnarray}
	\hwz(\vr, t)  & \equiv & \hwzp(\vr, t) + \hwzm(\vr, t), \label{eqn:defomega}\\
	\hwzp(\vr, t) & = & \sum_{i=1}^{N_+} \Omega \delta(\vr - \vr_i(t)), \label{eqn:point-vortex-p}\\
	\hwzm(\vr, t) & = & -\sum_{i=N_+ + 1}^{N_+ + N_-} \Omega \delta(\vr - \vr_i(t)),\label{eqn:point-vortex-m}
\end{eqnarray}
where $\vr=(x,y)$ is the position vector on $x-y$ plane, $\hwz(\vr,t)$ is the $z$-component of the vorticity, 
and $\delta(\vr)$ is the Dirac delta function in two dimensions.
The values of $N_+$ and $N_-$ are not necessarily the same.
The circulation of each point vortex is given by either $\Omega$ or $-\Omega$ where
$\Omega$ is a positive constant.
Magnitudes of $N_+\Omega$ and $N_-\Omega$ are finite.
The position vector of the $i$-th point vortex is given by
$\vr_i=\vr_i(t)$.
The discretized vorticities (\ref{eqn:point-vortex-p}) and (\ref{eqn:point-vortex-m}) are formal solutions of 
the 2D Euler equations (\ref{eqn:euler_p}) and (\ref{eqn:euler_m})
\begin{eqnarray}
	\frac{\partial }{\partial t} \hwzp(\vr, t) + \div (\hat{\vu}(\vr, t) \hwzp(\vr, t)) & = & 0, \label{eqn:euler_p}\\
	\frac{\partial }{\partial t} \hwzm(\vr, t) + \div (\hat{\vu}(\vr, t) \hwzm(\vr, t)) & = & 0 \label{eqn:euler_m}
\end{eqnarray}
where $\hat{\vu}(\vr,t)$ is the velocity field which is determined by the stream function $\hat{\psi}(\vr, t)$:
\begin{eqnarray}
	\hat{\vu}(\vr,t) & = & -\hat{\vec{z}} \times \grad \hat{\psi}, \\
	\hat{\psi}(\vr,t) & = & \int d\vr' G(\vr - \vr')\hwz(\vr',t)\nonumber\\
		& = & \sum_i\Omega_iG(\vr - \vr_i), \\
	G(\vr) & = & -\frac{1}{2\pi} \ln|\vr|.
\end{eqnarray}
Here, $\hat{\vec{z}}$ is the unit vector in the $z$-direction, 
and $G(\vr)$ is the 2D Green function for the Laplacian operator with an infinite domain.
From now on, for brevity, we shall omit the dependences on $t$ and $\vr$ and denote the two 
equations for the positive and the negative vortices into the single formula with double-sign,
if there is no ambiguity.
For example (\ref{eqn:euler_p}) and (\ref{eqn:euler_m}) are combined into the following form:
\begin{equation}
	\frac{\partial }{\partial t} \hwzpm + \div (\hat{\vu} \hwzpm) = 0. \label{eqn:euler_pm}
\end{equation}

As is mentioned in section \ref{sec:introduction}, we regard 
(\ref{eqn:euler_pm}) as the microscopic equations because they have 
the discretized point vortex solutions (\ref{eqn:point-vortex-p}) 
and (\ref{eqn:point-vortex-m}). 
Applying the Klimontovich formalism to (\ref{eqn:euler_pm}) \cite{Klimontovich}, 
we obtain an intermediate result which corresponds to (\ref{eqn:withcollision})
\begin{equation}
	\frac{\partial }{\partial t} \wzpm + \div(\vu \wzpm) = -\div \langle \delta\vu \delta \wzpm \rangle \label{eqn:collision-term-org}
\end{equation}
where the microscopic vorticity and the microscopic velocity field are defined by
\begin{eqnarray}
	\hwzpm 	& = & 		\langle\hwzpm\rangle     + \delta \wzpm	= \wzpm  + \delta \wzpm, \label{eqn:micro-w}\\
	\hat{\vec{u}} 	& = &		\langle \hat{\vu} \rangle + \delta \vec{u} = \vec{u} + \delta \vec{u}. \label{eqn:micro-u}
\end{eqnarray}
The term $\langle \delta\vu \delta \wzpm \rangle $ is a diffusion flux and will be denoted by $\bGamma_{\pm}$
\begin{eqnarray}
	\bGamma_{\pm} & \equiv & \langle \delta\vu\delta \wzpm\rangle \nonumber\\
		& = & -\int d\vr' \vec{F}(\vr - \vr') \langle \dwz' \delta \wzpm\rangle, \label{eqn:diffusion1}\\
	\vec{F}(\vr) & = & \hat{\vec{z}} \times \grad G(\vr), \\
	\delta \vec{u} & =& -\int d\vr' \vec{F}(\vr-\vr')\dwz'. \label{eqn:delta-u}
\end{eqnarray}
We note $\dwz'$ for $\dwz(\vr',t)$.
Similarly, we note $\wz'$ for $\wz(\vr',t)$.

To evaluate the diffusion fluxes $\bGamma_{\pm}$ explicitly, we 
introduce a small parameter $\epsilon$.
Orders are given by:
\begin{eqnarray}
	& & \wzpm \approx \grad^2\psi \approx O(\epsilon^0),\quad \vec{u} \approx \grad \psi \approx O(\epsilon^0),\quad \grad\vec{u} \approx \grad^2\psi \approx O(\epsilon^0),\nonumber\\
	& & \grad \wzpm \approx O(\epsilon^{1/2}),\quad \delta \wzpm \approx O(\epsilon^{1/2}),\quad \delta\vec{u} \approx O(\epsilon^{1/2}),\nonumber\\
	& & \frac{\partial \vec{u}}{\partial t} \approx O(\epsilon^{1/2}), \quad \frac{\partial \wzpm}{\partial t} \approx O(\epsilon^{1/2}), \quad \grad \grad \vec{u} \approx O(\epsilon^{1/2}),\nonumber\\
	& & \bGamma_{\pm} \approx O(\epsilon).
\end{eqnarray}
The expansion parameter $\epsilon$ is similar to the one introduced by Chavanis 
in \citeasnoun{Chavanis2001}, \citeasnoun{Chavanis2008}, \citeasnoun{Chavanis2012} 
and the references therein. 
In addition, we assume that the gradient of the vorticity
profile is weak.
Magnitude of $\Omega$ scales as either $1/N_+$ or $1/N_-$
and there is an upper limit of $N_{\pm}$.
These assumptions are due to the situation of the weak mean flow
and are necessary for the validity that a mean trajectory
is linear (\ref{eqn:integrated-linear-eq1}).
We will discuss the limitation on the number of vortices
in section \ref{subsec:epsilon}.
With these scalings, the left hand side of (\ref{eqn:collision-term-org})
is $O(\epsilon^{1/2})$, 
while the right hand side is $O(\epsilon^{3/2})$.
Expressing $\bGamma_{\pm}$ in the form of the perturbation expansion and gathering
the terms of the appropriate order, an analytical
formula for the diffusion fluxes will be obtained.

Although the detailed calculation process is not the same as the single-species case
presented in \citeasnoun{Yatsuyanagi2015}, there appears many similar techniques in the 
double species case.
Thus, the detailed process for deriving an explicit formula of the diffusion 
fluxes $\bGamma_{\pm}$ is given in appendix.

The final result is as follows.
As the obtained $\bGamma_{\pm}$ contain oscillatory terms,
we perform a space-average for $\bGamma_{\pm}$ to reveal
the characteristics of the collision term.
The space-averaged diffusion flux $\bGamma_s \equiv \langle \bGamma \rangle_s$ 
for the kinetic equation
\begin{equation}
	\frac{\partial \omega}{\partial t} + \div(\vu\omega) = -\div \bGamma_s(\vr), \label{eqn:euler-gamma}
\end{equation}
with 
\begin{equation}
	\wz = \wzp + \wzm
\end{equation}
is given by
\begin{eqnarray}
	\bGamma_s(\vr) & = & \bGamma_{s+}(\vr) + \bGamma_{s-}(\vr) \nonumber\\
	& = & -K \int d\vr' \frac{(\vu - \vu')(\vu-\vu')}{|\vu - \vu'|^3} \nonumber\\
	& & 	\cdot \left[ (\wzp' - \wzm') \grad \omega - (\wzp - \wzm) \grad'\omega' \right], \label{space-averaged-gamma-with-new-coeff}\\
	\bGamma_{s\pm}(\vr) & \equiv & -{\sf D}_s \cdot \grad \wzpm \pm \vec{V}_s \wzpm,\label{eqn:final-gamma}\\
	{\sf D}_s & = & K \int d\vr' \frac{(\vu - \vu')(\vu-\vu')(\wzp' - \wzm')}{|\vu - \vu'|^3}, \label{eqn:final-D}\\
	\vec{V}_s & = & K \int d\vr' \frac{(\vu - \vu')(\vu-\vu')\cdot \grad'\omega'}{|\vu - \vu'|^3}, \label{eqn:final-V} \\
	K & = & \frac{\Omega}{(2\pi)^3} \left( \frac{\pi}{L}\right)^2\frac{1}{k_{\rm min}},
\end{eqnarray}
where $K$ is a constant depending on $\Omega$, a system size $R$ and a coarse-graining 
scale $L$.
Parameter $k_{\rm min}$ is introduced to regularize a singularity,
and is determined by the largest wave length that does not exceed the system
size.
It is worth stressing that the term $\vec{V}_s\wzp$ has the opposite sign to the term $\vec{V}_s\wzm$.
It provides a mechanism for the ``charge separation" which is usually seen in the equilibrium distribution
for systems with positive and negative vortices \cite{Joyce,Yatsuyanagi2005}.
\section{Physical properties of the diffusion flux}
In this section, we examine several properties of the diffusion flux (\ref{space-averaged-gamma-with-new-coeff}).
\subsection{Diffusion flux in local and global equilibrium states}
At first, let us examine if the diffusion flux (\ref{space-averaged-gamma-with-new-coeff}) 
locally disappears in a local equilibrium state.
We rewrite (\ref{space-averaged-gamma-with-new-coeff}) into a symbolic form
\begin{equation}
	\bGamma_s(\vr) = - K \int d\vr' \vec{\gamma}[\omega, \psi; \omega', \psi'] \label{eqn:functional}
\end{equation}
where $\vec{\gamma}$ is a functional of $\omega$, $\psi$, $\omega'$ and $\psi'$.
Consider a state where temperature is locally uniform in each small region in the system,
namely there are small regions with different $\beta$.
In each small region, a local equilibrium condition is satisfied
\begin{equation}
	\omega_{{\rm leq}\pm} = \omega_{0\pm} \exp(\mp \beta_{\rm leq} \Omega\psi_{\rm leq}). \label{eqn:equilibrium}
\end{equation}
Inserting (\ref{eqn:equilibrium}) into $\vec{\gamma}$ in (\ref{eqn:functional})
and assuming that $\vr$ and $\vr'$ belong to the same subsystem,
we find that
\begin{eqnarray}
	& & \vec{\gamma}[\wz_{\rm leq},\psi_{\rm leq}; \wz'_{\rm leq}, \psi'_{\rm leq}]\nonumber\\
	& = & \frac{(\vu_{\rm leq} - \vu'_{\rm leq})(\vu_{\rm leq} - \vu'_{\rm leq})}{|\vu_{\rm leq} - \vu'_{\rm leq}|^3} \nonumber\\
	& & \cdot[ (\wz'_{{\rm leq}+} - \wz'_{{\rm leq}-}) \grad( \wz_{{\rm leq}+} + \wz_{{\rm leq}-})  - (\wz_{{\rm leq}+} - \wz_{{\rm leq}-})\grad'(\wz'_{{\rm leq}+} + \wz'_{{\rm leq}-})]\nonumber\\
	& = & -\beta_{\rm leq} \Omega (\wz'_{{\rm leq}+} - \wz'_{{\rm leq}-}) (\wz_{{\rm leq}+} - \wz_{{\rm leq}-}) \nonumber\\
	& &  \times \frac{(\vu_{\rm leq} - \vu'_{\rm leq})(\vu_{\rm leq} - \vu'_{\rm leq})}{|\vu_{\rm leq} - \vu'_{\rm leq}|^3} \cdot(\grad\psi_{\rm leq} - \grad'\psi'_{\rm leq})\nonumber\\
	& = & 0 \label{eqn:equilibrium2}
\end{eqnarray}
where $\vu_{\rm leq} = -\hat{\vec{z}} \times \grad \psi_{\rm leq}$ is used.
As $\vu_{\rm leq} - \vu'_{\rm leq}$ is perpendicular to $\grad\psi_{\rm leq} - \grad'\psi'_{\rm leq}$,
$\vec{\gamma}$ is equal to zero and this result indicates that the detailed balance
is achieved.

When the system reaches a global thermal equilibrium state 
characterized by
\begin{equation}
	\omega_{{\rm eq}\pm} = \omega_{0\pm} \exp(\mp \beta\Omega\psi_{\rm eq}), \label{eqn:global-equilibrium}\\
\end{equation}
with uniform $\beta$ \cite{Joyce},
we obtain
\begin{eqnarray}
	\grad'\wz'_{\rm eq} & = & \grad'(\wz'_{{\rm eq}+} + \wz'_{{\rm eq}+}) \nonumber\\
	& = & -\beta \Omega (\wz'_{{\rm eq}+} - \wz'_{{\rm eq}-})(\grad' \psi'_{\rm eq} - \grad \psi_{\rm eq} + \grad \psi_{\rm eq}).
\end{eqnarray}
As $(\vu_{\rm eq} - \vu'_{\rm eq}) \cdot (\grad'\psi'_{\rm eq} - \grad \psi_{\rm eq}) = 0$,
the drift term in (\ref{eqn:final-V}) is rewritten as
\begin{equation}
	\vec{V}_{s,{\rm eq}} = -\beta \Omega {\sf D}_{s,{\rm eq}} \cdot \grad\psi_{\rm eq} \label{eqn:Einsteinrel}
\end{equation}
and the total diffusion flux vanishes.
Equation (\ref{eqn:Einsteinrel}) is a counterpart of the Einstein relation \cite{Chavanis2001,Chavanis2008}.
\subsection{Sign of $d\wz/d\psi$ near thermal equilibrium states}
If the sign of the inverse temperature $\beta$ is negative, we obtain
\begin{equation}
	\frac{d\omega_{\rm eq}}{d\psi_{\rm eq}} = -\beta \Omega(\omega_{{\rm eq}+} - \omega_{{\rm eq}-}) \ge 0
\end{equation}
and
\begin{equation}
	\div (\vu_
{\rm eq} \omega_{\rm eq})  = 0
\end{equation}
where equations (\ref{eqn:global-equilibrium}) and
\begin{equation}
	\omega_{\rm eq} = \omega_{{\rm eq}+} + \omega_{{\rm eq}-}
\end{equation}
are used.
The point vortex system easily approaches a quasi-stationary state 
near the thermal equilibrium state by a violent relaxation
which is purely collisionless and driven by the mean field effects.
The local equilibrium state is also categorized in the above state.
In the local equilibrium state near the thermal equilibrium one,
the following relation is expected to be satisfied \cite{Yatsuyanagi2014}
\begin{equation}
	\div (\vec{u}_{\rm leq} \wz_{\rm leq}) \approx 0
\end{equation}
or equivalently
\begin{equation}
	\wz_{\rm leq} = \wz_{\rm leq}(\psi_{\rm leq}). \label{eqn:QSS}
\end{equation}
In this state, we may expect that
\begin{equation}
	\frac{d \omega_{\rm leq}}{d \psi_{\rm leq}} \ge 0\label{eqn:QSS2}
\end{equation}
almost everywhere in the system.
We will use this relation later.
\subsection{Energy-conservative property of diffusion flux}
It is shown that the obtained kinetic equation (\ref{eqn:euler-gamma})
conserves the total mean field energy $E$:
\begin{eqnarray}
	E & \equiv & \frac{1}{2} \int d\vr \psi \omega\nonumber\\
	& = & \frac{1}{2} \int d\vr \int d\vr' G(\vr-\vr') \omega'\omega.
\end{eqnarray}
Time derivative of the total mean field energy $E$ is given by
\begin{eqnarray}
	\frac{d E}{dt} & = & \frac{1}{2} \int d\vr \int d\vr'G(\vr - \vr')
		\left( \frac{\partial \omega'}{\partial t} \omega + \omega'\frac{\partial \omega}{\partial t}\right) \nonumber\\
	& = & \int d\vr\psi \frac{\partial \omega}{\partial t}, \label{eqn:time-derivative-system-energy}
\end{eqnarray}
Inserting the space-averaged equation of motion (\ref{eqn:euler-gamma})
into (\ref{eqn:time-derivative-system-energy}), we obtain
\begin{eqnarray}
	\frac{d E}{d t} & = & \int d\vr \psi \left( -\div (\vu \omega) - \div \bGamma_s \right) \nonumber\\
	& = & \int d\vr \grad \psi \cdot \vu\omega + \int d\vr \grad \psi \cdot \bGamma_s \nonumber\\
	& = & \int d\vr \grad \psi \cdot \bGamma_s \nonumber\\
	& = & - K \int d\vr \int d\vr'\grad \psi\cdot \frac{(\vu - \vu')(\vu - \vu')}{|\vu - \vu'|^3} \nonumber\\
	& & \cdot \left[ (\wzp' - \wzm')\grad \wz - (\wzp-\wzm) \grad'\wz'\right]. \label{eqn:dEdt1}
\end{eqnarray}
By permuting the dummy variables $\vr$ and $\vr'$ in (\ref{eqn:dEdt1})
and taking the half-sum of the resulting expressions, we obtain
\begin{eqnarray}
	\frac{d E}{dt} & = & - \frac{K}{2}\int d\vr \int d\vr'(\grad \psi - \grad'\psi')\cdot \frac{\vu - \vu'}{|\vu - \vu'|^3} \nonumber\\
	& & \times (\vu - \vu')\cdot \left[ (\wzp' - \wzm')\grad \wz - (\wzp-\wzm) \grad'\wz'\right]\nonumber\\
	& = & 0. \label{eqn:dEdt2}
\end{eqnarray}
We conclude that the obtained diffusion flux conserves the total mean field energy.

It is also revealed that the energy conservation is achieved by ballancing the energy dissipation process due to
the diffusion term and the energy production process due to the drift
term.
We divide the expression (\ref{eqn:dEdt1}) into two parts, namely the term
which corresponds to the diffusion term and the one to the drift term.
\begin{eqnarray}
	\frac{dE}{dt} & = & \left.\frac{dE}{dt}\right|_{\sf D} + \left.\frac{dE}{dt}\right|_{\vec{V}} = 0,\\
	\left.\frac{dE}{dt}\right|_{\sf D} & = & - K \int d\vr \int d\vr'\grad \psi\cdot \frac{(\vu - \vu')(\vu - \vu')}{|\vu - \vu'|^3} 
		\cdot (\wzp' - \wzm')\grad \wz \label{eqn:dedt_d1},\\
	\left.\frac{dE}{dt}\right|_{\vec{V}} & = & K \int d\vr \int d\vr'\grad \psi\cdot \frac{(\vu - \vu')(\vu - \vu')}{|\vu - \vu'|^3} 
		\cdot (\wzp - \wzm) \grad'\wz'\label{eqn:dedt_v1}.
\end{eqnarray}
If the vorticity $\omega$ is a function of the stream function $\psi$ (see (\ref{eqn:QSS})),
equations (\ref{eqn:dedt_d1}) and (\ref{eqn:dedt_v1}) are rewritten as
\begin{eqnarray}
	\left.\frac{dE}{dt}\right|_{\sf D} & = & - K \int d\vr \int d\vr' \frac{|\grad \psi\cdot(\vu - \vu')|^2}{|\vu - \vu'|^3} 
		(\wzp' - \wzm')\frac{d \omega}{d \psi} ,\\
	\left.\frac{dE}{dt}\right|_{\vec{V}} & = & K \int d\vr \int d\vr' \frac{|\grad \psi\cdot (\vu - \vu')|^2}{|\vu - \vu'|^3} 
		(\wzp - \wzm)\frac{d \omega'}{d \psi'} .
\end{eqnarray}
Thus, if $d \omega/d \psi \ge 0$ with $\beta < 0$, it is concluded that
\begin{equation}
	\left.\frac{dE}{dt}\right|_{\vec{V}} = - \left.\frac{dE}{dt}\right|_{\sf D} \ge 0 ,\label{eqn:energy-balance}
\end{equation}
namely, 
\begin{equation}
	\left.\frac{dE}{dt}\right|_{\vec{V}} + \left.\frac{dE}{dt}\right|_{\sf D} = 0 .
\end{equation}
\subsection{H theorem\label{sec:htheorem}}
It is shown that the obtained kinetic equation (\ref{eqn:euler-gamma})
satisfies an $H$ theorem.
The entropy function $S$ is defined by using the $H$ function:
\begin{eqnarray}
	S & = & -k_B H,\label{eqn:entropy}\\
	H & = & \int d\vr\frac{\wzp}{\Omega}\ln \frac{\wzp}{\Omega} + \frac{\wzm}{-\Omega}\ln \frac{\wzm}{-\Omega} + {\rm const.}\nonumber\\
	& = & \frac{1}{\Omega}\int d\vr \wzp \ln \wzp - \wzm \ln |\wzm| - 2N\ln\Omega + {\rm const. }
\end{eqnarray}
The time derivative of the $H$ function is given by
\begin{eqnarray}
	\frac{dH}{dt} & = & \frac{1}{\Omega} \int d\vr \frac{\partial \wzp}{\partial t} (\ln \wzp + 1) - \frac{\partial \wzm}{\partial t} (\ln |\wzm| + 1)\nonumber\\
	& = & \frac{1}{\Omega} \int d\vr \vu \wzp \cdot \grad \ln \wzp - \vu \wzm \cdot \grad \ln |\wzm|\nonumber\\
	& & + \frac{1}{\Omega} \int d\vr \bGamma_{s+} \cdot \grad \ln \wzp - \bGamma_{s-} \cdot \grad \ln|\wzm|\nonumber\\
	& = & \frac{1}{\Omega} \int d\vr \vu \cdot \grad \wzp - \vu \cdot \grad \wzm\nonumber\\
	& & + \frac{1}{\Omega} \int d\vr \bGamma_{s+}\cdot\grad\ln \wzp - \bGamma_{s-}\cdot\grad \ln |\wzm|\nonumber\\
	& = & \frac{1}{\Omega} \int d\vr \bGamma_{s+}\cdot\grad\ln \wzp - \bGamma_{s-}\cdot\grad \ln |\wzm|. \label{eqn:dhdt0}
\end{eqnarray}
For simplicity, we introduce the following notations.
\begin{eqnarray}
	& & \grad \ln \wzp = \vec{v}_+,\quad\grad \ln |\wzm| = \vec{v}_-,\nonumber\\
	& & \vu - \vu' = \vec{U}.
\end{eqnarray}

Inserting (\ref{space-averaged-gamma-with-new-coeff}) into (\ref{eqn:dhdt0}),
we obtain
\begin{eqnarray}
	\frac{dH}{dt} & = & - \frac{K}{\Omega} \int d\vr \int d\vr' \vec{v}_+ \cdot \frac{\vec{U}\vec{U}}{|\vec{U}|^3}\cdot [(\wzp' - \wzm') \grad\wzp - \wzp\grad'\wz'] \nonumber\\
		& & - \frac{K}{\Omega} \int d\vr \int d\vr' \vec{v}_- \cdot \frac{\vec{U}\vec{U}}{|\vec{U}|^3}\cdot [(\wzp' - \wzm') \grad\wzm + \wzm\grad'\wz']\nonumber\\
	& = & - \frac{K}{\Omega} \int d\vr \int d\vr' \frac{1}{|\vec{U}^3|} \nonumber\\
	& & \times \left[ 
		\wzp\wzp'(\vec{v}_+\cdot\vec{U}\vec{U}\cdot(\vec{v}_+ - \vec{v}'_+)
	\right.\nonumber\\
		& & + 		\wzm\wzm'(\vec{v}_-\cdot\vec{U}\vec{U}\cdot(\vec{v}_- - \vec{v}'_-)\nonumber\\
		& & - 		\wzp\wzm'(\vec{v}_+\cdot\vec{U}\vec{U}\cdot(\vec{v}_+ + \vec{v}'_-)\nonumber\\
		& & -	\left.
		\wzm\wzp'(\vec{v}_-\cdot\vec{U}\vec{U}\cdot(\vec{v}_- + \vec{v}'_+)
	\right].  \label{eqn:dHdt1}
\end{eqnarray}
By permuting the dummy variables $\vr$ and $\vr'$ in (\ref{eqn:dHdt1})
and taking the half-sum of the resulting expressions, we obtain
\begin{eqnarray}
	\frac{dH}{dt} & = & - \frac{K}{2 \Omega} \int d\vr \int d\vr' \frac{1}{|\vec{U}|^3}\nonumber\\
	& & \times [
		\wzp\wzp'|(\vec{v}_+ - \vec{v}'_+)\cdot \vec{U}|^2 \nonumber\\
	& & +   \wzm\wzm'|(\vec{v}_- - \vec{v}'_-)\cdot \vec{U}|^2 \nonumber\\
	& & -   \wzp\wzm'|(\vec{v}_+ - \vec{v}'_-)\cdot \vec{U}|^2 \nonumber\\
	& & -   \wzm\wzp'|(\vec{v}_- - \vec{v}'_+)\cdot \vec{U}|^2]\le 0. \label{eqn:dHdt2}
\end{eqnarray}
As the integrand of (\ref{eqn:dHdt2}) is positive or equal to zero,
$dH/dt$ is negative or equal to zero.
It is concluded that the entropy function (\ref{eqn:entropy})
is the monotonically increasing function.

It is also revealed that the diffusion term increases the entropy, while the drift term decreases it.
We divide the expression (\ref{eqn:dhdt0}) into two parts (the entropy function $S$ is 
used instead of the $H$ function).
\begin{eqnarray}
	\frac{dS}{dt} & = & \left.\frac{dS}{dt}\right|_{\sf D} + \left.\frac{dS}{dt}\right|_{\vec{V}} \ge 0 ,\\
	\left.\frac{dS}{dt}\right|_{\sf D} & = & \frac{k_B}{\Omega} \int d\vr \frac{\grad \wzp \cdot {\sf D}_s \cdot \grad \wzp}{\wzp} - \frac{\grad \wzm \cdot {\sf D}_s \cdot \grad \wzm}{\wzm} \nonumber\\
		& = & k_B \frac{K}{\Omega} \int d\vr \int d\vr' \left(\frac{|\grad \wzp \cdot (\vu - \vu')|^2}{\wzp} - \frac{|\grad \wzm \cdot (\vu - \vu')|^2}{\wzm} \label{eqn:dhdt_d1} \right)\nonumber\\
		& & \quad	\times \frac{\wzp'- \wzm'}{|\vu - \vu'|^3},\\
	\left.\frac{dS}{dt}\right|_{\vec{V}} & = & -\frac{k_B}{\Omega} \int d\vr \vec{V}_s \cdot \grad \wz\nonumber\\
		& = & - k_B \frac{K}{\Omega} \int d\vr \int d\vr'\grad \wz \cdot \frac{(\vu-\vu')(\vu-\vu')}{|\vu-\vu'|^3}\cdot\grad'\wz'\label{eqn:dhdt_v1} .
\end{eqnarray}
Equation (\ref{eqn:dhdt_d1}) indicates that
\begin{equation}
	\left.\frac{dS}{dt}\right|_{\sf D} \ge 0
\end{equation}
regardless of the sign of $d \wz/ d \psi$.
On the other hand, if the system reaches a local equilibrium state,
we expect that (\ref{eqn:QSS2}) is valid.
In this case, equation (\ref{eqn:dhdt_v1}) is rewritten as
\begin{equation}
	\left.\frac{dS}{dt}\right|_{\vec{V}} = -k_B \frac{K}{\Omega} \int d\vr \int d\vr'\frac{|\grad \psi \cdot (\vu - \vu')|^2}{|\vu - \vu'|^3} \frac{d \wz}{d \psi} \frac{d \wz'}{d \psi'}
\end{equation}
and we obtain 
\begin{equation}
	\left.\frac{dS}{dt}\right|_{\vec{V}} \le 0
\end{equation}
in the local equilibrium state.
When the system reaches a thermal equilibrium state,
the relation
\begin{equation}
	\left.\frac{dS}{dt}\right|_{\vec{V}} = - \left.\frac{dS}{dt}\right|_{\sf D} \le 0
\end{equation}
namely,
\begin{equation}
	\left.\frac{dS}{dt}\right|_{\vec{V}} + \left.\frac{dS}{dt}\right|_{\sf D} = 0
\end{equation}
is satisfied, and the entropy production stops.

This clearly indicates the crucial role of the drift term in the self-organization of 
the point vortex system.
The diffusion term increases the entropy, while the drift term decreases it.
Namely, in the self-organization of the 2D point vortex system with negative $\beta$,
a background distribution outside the clumps is necessary to dump the entropy.
The clump formation is driven by the drift term and the background distribution
outside the clumps is made by the diffusion term.
This conclusion is supported by the nonneutral plasma experiments \cite{Sanpei,Soga,Jin,Fine}
and the numerical simulation \cite{Yatsuyanagi2005}.
\section{Numerical result}
\begin{figure}
\begin{center}
\resizebox{10cm}{!}{\includegraphics{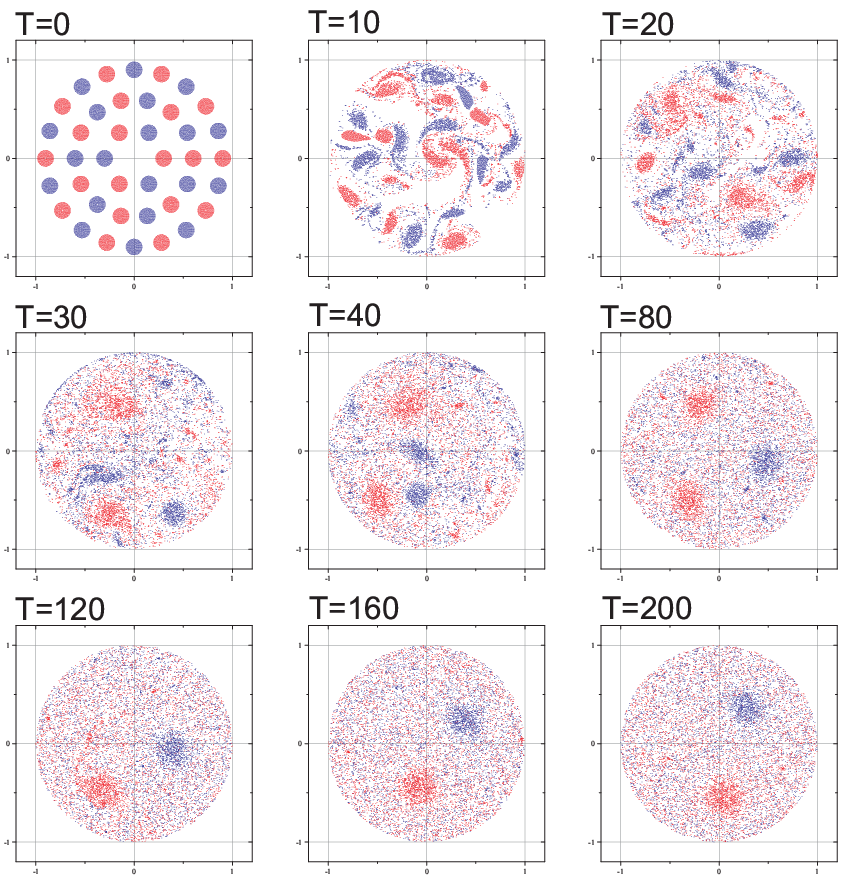}}
\caption{Time evolution of the vortices up to 40 turnover time
of an initial small clump is shown. 
Characteristic turnover time
is given by $T \approx 5$.
Initial profile is quickly destroyed and two relatively
large clumps with same-sign vortices are formed by the self-organizing
feature of the system.
}
\label{fig:simulation}
\end{center}
\end{figure}

We will demonstrate an example of the self-organization in the 2D
point vortex system by numerical simulations.
Twenty positive (red) and 20 negative (blue) clumps are initially arranged 
uniformly in a circular wall with radius $R$.
Each clump is composed of the same-sign 283 vortices.
Total number of the vortices is $283\times40=11320$.
Characteristic time scale is given by a self-rotation 
time of a small clump $T \approx 5$.
A motion of the vortices are traced by the following equation
of motion
\begin{equation}
	\frac{\partial \vr_i}{\partial t} = \frac{1}{2\pi}\sum_{j \neq i} \Omega_i \frac{\hat{\vec{z}} \times (\vr_i - \vr_j)}{|\vr_i-\vr_j|^2}
		- \frac{1}{2\pi}\sum_{j } \Omega_i \frac{\hat{\vec{z}} \times (\vr_i - \bar{\vr}_j)}{|\vr_i-\bar{\vr}_j|^2}
\end{equation}
where the circular wall effect is introduced by image vortices located at
\begin{equation}
	\bar{\vr}_i = \frac{R^2}{|\vr_i|^2} \vr_i.
\end{equation}
Time evolution of the system is given in figure \ref{fig:simulation}.
The system finally settles down into a well-known 
self-organized thermal equilibrium state which is 
described by the sinh-Poisson equation \cite{Joyce,Pointin1976}.
Remarkable features of this result are the charge-separation
of the vortices and the condensation of the same-sign vortices.
Energy belonging to the $i$-th point vortex $H_i$ is defined by
\begin{eqnarray}
	H & = & \sum_i H_i, \nonumber\\
	H_i & = & \frac{1}{2} \Omega_i \psi_i,\nonumber\\
	\psi_i & = & -\frac{1}{2\pi} \sum_{j \neq i} \Omega_j \ln|\vr_i - \vr_j|
	+ \frac{1}{2\pi} \sum_{j} \Omega_j \ln|\vr_i - \bar{\vr}_j|\nonumber\\
	& & \qquad - \frac{1}{2\pi} \sum_{j} \Omega_j \ln\left|\frac{R}{\left|\vr_j\right|}\right|
\end{eqnarray}
where $H$ is a total energy of the system.
Energy of the vortices inside a clump is positive and large
as the same-sign vortices are confined in a small region.
Such configuration is enabled by the drift term
as is shown in \eref{eqn:energy-balance}.
In addition, the drift term plays an important role of separation
of the positive and negative vortices.
The positive vortices and the negative vortices are driven
in the opposite directions by the drift term as the signs of the
drift terms for the positive and the negative vortices
in \eref{eqn:final-gamma} are opposite.

On the other hand, the system is an energy-conserving one.
Thus, if there are vortices that gains energy by the clumping,
there must be vortices that loses energy.
These vortices that loses energy go outside a clump and form 
a background distribution.
This feature to lower the energy of the vortices is provided
by the diffusion term as is shown in \eref{eqn:energy-balance}.

\begin{figure}
\begin{center}
\resizebox{10cm}{!}{\includegraphics{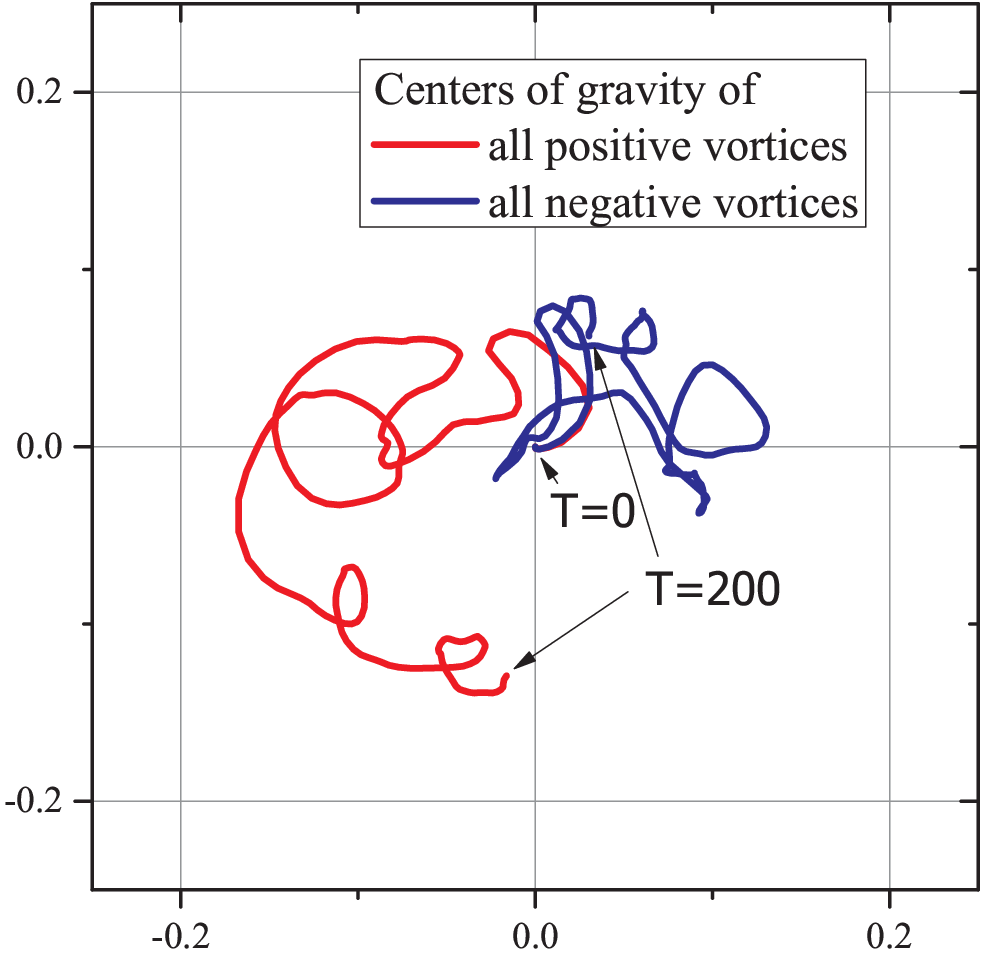}}
\caption{
Trajectories of centers of gravity of the positive (red line) and 
negative (blue line) vortices are plotted.
Length scale is normalized by the radius of the circular wall $R$.
Initial positions of both centers of gravity locate at the center
of the wall.
The distance between the both centers of gravity at the end of the simulation
is approximately $0.20R$.
}
\label{fig:charge-separation}
\end{center}
\end{figure}

Another evidence for the charge-separation is given in 
figure \ref{fig:charge-separation}.
It can be seen that the center of gravity of the positive vortices
goes downward and the center of gravity of the negative vortices
goes upward.
This figure clearly indicates the charge separation.

Following these observations, we conclude that Fokker-Planck
type collision term for double-species 2D point vortex system provides 
an essential and crucial role for the self-organization in the system
at negative absolute temperature.
\section{Discussion\label{sec:discussion}}
\subsection{Confinment of positive and negative vortices in an infinite domain}
In our kinetic theory, the vortices are located in an infinite domain 
without any boundary. A positive and a negative vortices 
with the same strength move in parallel straight lines. Many pairs of 
the positive and negative vortices escape from the initial center of 
vortices without violating the conservation of the inertia $I$,
\begin{equation}
	I = \sum_i \Omega_i |\vr_i|^2.\label{eqn:inertia}
\end{equation}
However, if the system temperature is negative, clumps consisting of the same 
sign vortices are expected to be formed by the rapid violent relaxation 
and two clumps with the different signs travel along the perpendicular bisector of the line 
joining the two centers of the clumps with nearly constant speed. So 
even if there is no boundary, our kinetic theory describes the relaxation 
process in the reference flame moving with the center of the two clumps.
\subsection{Magnitude of the expansion parameter $\epsilon$ \label{subsec:epsilon}}
We have estimated the magnitude of the expansion parameter
$\epsilon$ in the previous paper \cite{Yatsuyanagi2015}.
We find a new estimation and will present the result.

Let us reexamine the order of $\epsilon$, which is given by 
the ratio of the drift velocity to the macroscopic 
fluid velocity,
\begin{eqnarray}
	\frac{|\vec{V}|}{|\vec{u}|} & = & \frac{\left|K \int d\vr' \frac{(\vu - \vu')(\vu-\vu')\cdot \grad'\omega'}{|\vu - \vu'|^3}\right|}{\left|\vec{u}\right|} \nonumber\\
	& = & O\left(\frac{1}{16} \frac{1}{N_{\pm}} \left(\frac{R}{L}\right)^2\right).
\end{eqnarray}
Here, we have used the relations $k_{\rm min} = 2\pi/R$, $|\vec{u}| = R\omega$
and $\omega=N_{\pm}\Omega/(\pi R^2)$, which are introduced in 
\citeasnoun{Yatsuyanagi2015}. The notation $R$ is the characteristic length of 
the system, $L (< R)$ the space-averaging size introduced in Appendix. 
As the order of the obtained diffusion flux (\ref{space-averaged-gamma-with-new-coeff}) 
is $O(\epsilon)$, the following scaling is obtained,
\begin{equation}
	\epsilon \approx \frac{1}{16}\frac{1}{N_{\pm}} \left(\frac{R}{L}\right)^2.
\end{equation}
Let us introduce a notation $N_L$ as
\begin{equation}
	N_L \equiv N_{\pm} \left( \frac{L}{R}\right)^2.
\end{equation}
This notation represents the number of vortices inside
the space-averaging (coarse-graining) area with sides $L \times L$.
Finally, the smallness parameter $\epsilon$ is characterized by
\begin{equation}
	\epsilon \approx \frac{1}{N_L} > \frac{1}{N_{\pm}}.
\end{equation}
where $N_{\pm}$ is given by a counterpart of \eref{eqn:limit}
\begin{equation}
	1 < \pi \left(\frac{R}{L}\right)^2 < N_{\pm} < \frac{\pi}{16}\left(\frac{R}{L}\right)^4.
\end{equation}
This quantity directly corresponds to the discreteness of matter
for our kinetic theory.
\subsection{Comparison of the directions of the diffusion and the drift
with an ordinary Fokker-Planck system}
In an ordinary Fokker-Planck system with positive $\beta$, particles are 
populated in a low energy state and the diffusion occurs toward
a high energy state.
Low energy particles diffuse toward the high energy state by the diffusion,
while high energy particles lose their energy by the friction and
go to the low energy state.
On the other hand, in a point vortex system with negative
$\beta$, vortices are populated in a high energy state and the diffusion
occurs toward a low energy state.
High energy vortices diffuse toward the low energy state by the diffusion,
while low energy vortices gain their energy by the drift and go
to the high energy state.
Namely, regardless of the sign of $\beta$, the diffusion term works
to lower the population of the particles (vortices). 
On the other hand, when $\beta > 0$, the friction
term decreases the speed of the diffusion. 
When $\beta < 0$, the drift term works to accumulate the vortices
against the diffusion.
This effect of the drift term may be called ``negative friction".
Note that the directions of the diffusive effect and the drift effect
are always opposite regardless the sign of $\beta$.
\begin{figure}
\begin{center}
\resizebox{10cm}{!}{\includegraphics{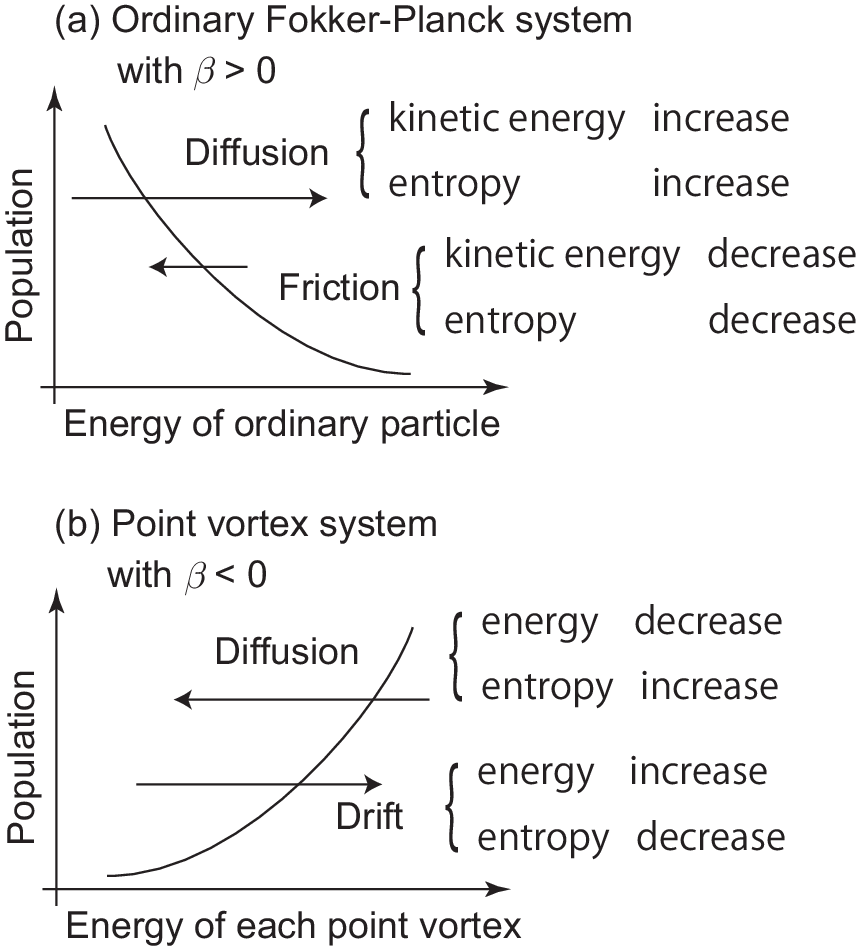}}
\caption{Populations in the energy space are illustrated.
In an ordinary Fokker-Planck system with positive $\beta$, particles are 
populated in a low energy state and the diffusion occurs toward
a high energy state.
On the other hand in a point vortex system with negative
$\beta$, vortices are populated in a high energy state and the diffusion
occurs toward a low energy state.
}
\label{fig:time-evolution}
\end{center}
\end{figure}
\section{Conclusion}
We have demonstrated the simple and explicit formula (\ref{space-averaged-gamma-with-new-coeff})
of the Fokker-Planck type collision term for double-species point vortex system
without the collective effect.
We have also demonstrated the strong evidence of the important role of the drift
term in the self-organization of the 2D point vortex system at negative absolute 
temperature.

The previous model for the single-species point vortex system
\cite{Yatsuyanagi2015} corresponds to the guiding-center nonneutral plasmas.
However, as the nonneutral plasma is not a perfect tool to understand all the
phenomena in the 2D turbulence,
we have motivated to extend the previous result 
to the double-species system allowing the different 
numbers of the positive and the negative vortices.
The current model can be applied to researches on the 2D turbulence,
including plasmas with the same magnitudes of the charges and the masses, 
e.g., electron-positron plasmas and hydrogen-antihydrogen plasmas.

The obtained diffusion flux $\bGamma_s$ conserves the mean field energy.
The $H$ theorem ensures that a point vortex system of any type of the flow, 
including an axisymmetric one, relaxes to the Boltzmann thermal
equilibrium state (\ref{eqn:global-equilibrium}).
The positive and the negative vortices independently relax to the thermal equilibrium states
even if the numbers of the positive and the negative vortices are different.

It should be noted again that the drift term plays the important role in the
self-organization in the 2D point vortex system.
In (\ref{eqn:final-gamma}), the sign of the drift term changes in accordance with the 
sign of the vorticity, while the sign of the diffusion term is always negative.
This implies that the drift term provides the ``charge separation" of the
vortices, which is commonly observed in equilibrium states at negative 
temperature.

During a relaxation process in a closed system, (Boltzmann) entropy
should increase even if the system energy is conserved.
It is reasonable that a distribution of particles broaden and finally reaches 
a flat distribution.
Thus, it is difficult to understand the clump formation 
usually seen in the 2D self-organization as the entropy seems to decrease.
As is discussed in section \ref{sec:htheorem}, it is found that the drift term
decreases the entropy, while the total entropy increases.
The effect of the drift term to decrease the entropy, say negative
entropy production, hide 
behind the effect of the diffusion term to increase the entropy.
It has also been stressed the common and essential role of the background 
vortices in supporting the vortex condensation experimentally 
\cite{Sanpei,Soga,Jin,Fine} and numerically \cite{Yatsuyanagi2005}.
This role is provided by the drift term.
Note that the negative entropy production and the negative friction
exist also in a single-species point vortex system.

It was reported that a sinh-Poisson equilibrium state is observed
in a 2D Navier-Stokes system with finite Reynolds number \cite{Montgomery1993,Li1996,Li1997},
although the sinh-Poisson equation is derived not in a continuous fluid system 
but in a discretized point vortex system \cite{Joyce}.
It is conjectured that the collision term in the Navier-Stokes equation 
implicitly involves a turbulent drift-like effect at high Reynolds number
in addition to the turbulent diffusion.

There are several outstanding issues remaining.
First, the final formulae (\ref{eqn:final-D}) and (\ref{eqn:final-V}) include
unknown parameters $k_{min}$ and $L$.
Second, the integrals in (\ref{eqn:final-D}) for ${\sf D}_s$
and (\ref{eqn:final-V}) for $\vec{V}_s$ 
contain the divergent integrand, although combined terms
$\bGamma_{s} = \bGamma_{s+} + \bGamma_{s-}$
are regularized.
A method to resolve this problem may be to introduce a corrective effect,
some kind of screening.
Third, the obtained diffusion flux does not conserve the inertia (\ref{eqn:inertia})
for an arbitrary flow.
If the flow is axisymmetric, the inertia conserves.
A more rigorous justification will be needed for fixing the above issues.
\ack
The authors are grateful to Prof. P. H. Chavanis for helpful comments.
This work was supported by JSPS KAKENHI Grant Number 24540400.

\appendix
\section{Outline of the calculation}
In the following, we show an outline for deriving an explicit formula for the diffusion 
fluxes $\bGamma_{\pm}$.

To rewrite the diffusion fluxes (\ref{eqn:collision-term-org}),
we introduce linearized equations obtained by inserting 
(\ref{eqn:micro-w}) and (\ref{eqn:micro-u})
into  (\ref{eqn:euler_pm}) and assembling the terms of the order up to $\epsilon$ \cite{Yatsuyanagi2015}:
\begin{equation}
	\frac{\partial }{\partial t} \delta \wzpm + \div(\vu\delta\wzpm) = -\delta\vu \cdot \grad \wzpm. \label{eqn:linearlized}
\end{equation}
As the macroscopic quantities $\vec{u}$ appearing in the second term in the 
left-hand side and $\grad \wzpm$ in the right-hand side are supposed to be 
constant in the time scale of the microscopic fluctuation,
equations (\ref{eqn:linearlized}) can be integrated
\begin{eqnarray}
	\delta \wzpm & = & -\int_{t_0}^t d\tau \delta\vu(\vr - \vu(t-\tau),\tau) \cdot \grad \wzpm \nonumber\\
		& & + \delta\wzpm(\vr - \vu(t - t_0),t_0). \label{eqn:integrated-linear-eq1}
\end{eqnarray}
This approximation means that the mean trajectory is linear (straight)
\cite{Chavanis2008,Yatsuyanagi2015}.
The value of $t_0$ is chosen to satisfy $t - t_0 \gg t_c$ where $t_c$ is a 
correlation time of the fluctuation.

Substituting (\ref{eqn:integrated-linear-eq1}) into the correlation terms in 
(\ref{eqn:diffusion1}), we obtain
\begin{eqnarray}
	& & \bGamma_{\pm} \nonumber\\
	& = & -\int_{t_0}^t d\tau\int d\vr'\int d\vr''\vec{F}(\vr-\vr') \vec{F}(\vr - \vu(t-\tau) - \vr'') \cdot \grad \wzpm \nonumber\\
	& & 	\times \left\langle \dwz(\vr'',\tau) \dwz'\right\rangle\nonumber\\
	& & -\int_{t_0}^t d\tau\int d\vr'\int d\vr''\vec{F}(\vr-\vr') \vec{F}(\vr' - \vu'(t-\tau) - \vr'') \cdot \grad' \wz' \nonumber\\
	& & 	\times \left\langle \dwz(\vr'',\tau) \delta\wzpm\right\rangle \label{eqn:diffusion-friction}
\end{eqnarray}
where $\grad'\wz' = \grad_{\vr'}\wz(\vr')$.
There are three correlation terms $\left\langle \dwz(\vr'',\tau) \dwz'\right\rangle$, $\left\langle \dwz(\vr'',\tau) \delta\wzp\right\rangle$
and $\left\langle \dwz(\vr'',\tau) \delta\wzm\right\rangle$ in (\ref{eqn:diffusion-friction}).
At first, we handle the term $\left\langle \dwz(\vr'',\tau) \delta\wzp\right\rangle$.
\begin{eqnarray}
	& & \left\langle \dwz(\vr'',\tau) \delta\wzp\right\rangle \nonumber\\
	& = & \left\langle \sum_{i=1}^{N_+ } \Omega_i^2 \delta(\vr''-\vr_i(\tau))\delta(\vr-\vr_i)\right\rangle\nonumber\\
	& & + \left\langle \sum_{i=1}^{N_+ + N_-} \sum_{j\neq i}^{N_+ }\Omega_i\Omega_j\delta(\vr''-\vr_i(\tau)) \delta(\vr-\vr_j)\right\rangle\nonumber\\
	& & - \wz(\vr'',\tau) \wzp .\label{eqn:diffusion6}
\end{eqnarray}
The first term in the right-hand side in (\ref{eqn:diffusion6}) corresponds
to the case of $i=j$,
and the second term corresponds to the case of $i\neq j$.

For the $i=j$ case, the formula is rewritten as
\begin{eqnarray}
	& & \left\langle \sum_{i=1}^{N_+} \Omega_i^2 \delta(\vr'' - \vr_i(\tau) - \vr + \vr_i) \delta(\vr-\vr_i)\right\rangle\nonumber\\
	& = & \sum_{i=1}^{N_+ } \Omega_i^2 \left\langle \delta(\vr''-\vr + \vu_i(t-\tau) + \bxi_i)\right\rangle_{\xi}\left\langle \delta(\vr - \vr_i)\right\rangle\nonumber\\
	& = & \left\langle \delta(\vr''-\vr + \vu(t-\tau)+ \bxi)\right\rangle_{\xi}\Omega \wzp
\end{eqnarray}
Here we introduce a stochastic process to evaluate $\vr_i - \vr_i(\tau)$
\begin{eqnarray}
	\vr_i-\vr_i(\tau) & = & \int_{\tau}^t\vu(\vr_{i}(\tau'),\tau')d\tau' + \bxi_i \nonumber\\
	& \approx & \vu_{i}(t-\tau) + \bxi_i.
\label{eqn:diffusion8}
\end{eqnarray}
The first term in (\ref{eqn:diffusion8}) represents the
approximation that the mean trajectory is linear and the second term 
represents a Brownian motion.
The stochastic process represented by $\langle \cdot \rangle_{\xi}$ includes all the possible motions
to reach position $\vr_i$ at time $t$.

For the $i \neq j$ case, we introduce an approximation that the correlation between the particles 
can be neglected and the term is rewritten as
\begin{eqnarray}
	& & \left\langle \sum_{i=1}^{N_+ + N_-} \sum_{j\neq i}^{N_+ }\Omega_i\Omega_j\delta(\vr''-\vr_i(\tau)) \delta(\vr-\vr_j)\right\rangle\nonumber\\
	& = & \left( \wzp(\vr'',\tau) + \wzm(\vr'',\tau) \right) \wzp \nonumber\\
	& & - \frac{1}{N_+}\wzp(\vr'',\tau)\wzp ,
\end{eqnarray}
assuming
\begin{eqnarray}
	\sum_{i=1}^{N_+ + N_-}\Omega_i\left\langle \delta(\vr''-\vr_i(\tau))\right\rangle & = & \wzp(\vr'',\tau) + \wzm(\vr'',\tau),\\
	\wzp(\vr'',\tau) & = & N_+ \Omega\left\langle \delta(\vr''-\vr_i(\tau))\right\rangle,\label{eqn:approx2a}\\
	\wzm(\vr'',\tau) & = & -N_- \Omega\left\langle \delta(\vr''-\vr_i(\tau))\right\rangle.\label{eqn:approx2b}
\end{eqnarray}
Combining the results of the $i=j$ and the $i\neq j$ cases, we rewrite (\ref{eqn:diffusion6}) as
\begin{eqnarray}
	& & \left\langle \dwz(\vr'',\tau) \dwz_{+} \right\rangle\nonumber\\
	& = & \Omega \left\langle \delta(\vr'' - \vr + \vu(t-\tau) + \vec{\xi}) \right\rangle_{\xi} \wzp \nonumber\\
	& & - \frac{1}{N_{+}} \wzp(\vr'',\tau) \wzp. \label{eqn:diffusion10a}
\end{eqnarray}
Similarly, we obtain
\begin{eqnarray}
	& & \left\langle \dwz(\vr'',\tau) \dwz_{-} \right\rangle\nonumber\\
	& = & - \Omega \left\langle \delta(\vr'' - \vr + \vu(t-\tau) + \vec{\xi}) \right\rangle_{\xi} \wzm \nonumber\\
	& & - \frac{1}{N_{-}} \wzm(\vr'',\tau) \wzm. \label{eqn:diffusion10b}
\end{eqnarray}
\begin{eqnarray}
	& & \left\langle \dwz(\vr'',\tau) \dwz'\right\rangle\nonumber\\
	& = & \Omega \left\langle \delta(\vr'' - \vr' + \vu'(t-\tau) + \vec{\xi}) \right\rangle_{\xi} \left( \wzp' - \wzm' \right) \nonumber\\
	& & - \frac{1}{N_+} \wzp(\vr'',\tau) \wzp' - \frac{1}{N_-} \wzm(\vr'',\tau) \wzm'). \label{eqn:diffusion11}
\end{eqnarray}

To proceed with the evaluation of (\ref{eqn:diffusion10a}), (\ref{eqn:diffusion10b}) and (\ref{eqn:diffusion11}), conservation laws are introduced
\begin{eqnarray}
	\int d\vr' \left\langle \dwz(\vr'',\tau) \dwz'\right\rangle & = & 0, \label{eqn:conservation-law1}\\
	\int d\vr \left\langle\dwz(\vr'',\tau) \delta \wzpm(\vr,t) \right\rangle & = & 0. \label{eqn:conservation-law2}
\end{eqnarray}
Using these formulae, each term in (\ref{eqn:diffusion-friction})
is evaluated.
\begin{eqnarray}
	& & -\int_{t_0}^t d\tau\int d\vr'\int d\vr''\vec{F}(\vr-\vr') \vec{F}(\vr - \vu(t-\tau) - \vr'') \cdot \grad \wzpm \nonumber\\
	& & 	\times \left\langle \dwz(\vr'',\tau) \dwz'\right\rangle\nonumber\\
	& = & -\Omega\int d\vr'\vec{F}(\vr - \vr')\int \frac{d\vk}{(2\pi)^2}\nonumber\\
	& & \times \exp(i\vk\cdot(\vr - \vr')) \frac{\hat{\vec{z}}\times i\vk}{|\vk|^2}\cdot \left( \wzp' - \wzm' \right) \grad \wzpm \nonumber\\
	& & \times\left[\pi\delta(\vk \cdot (\vu - \vu')) - \frac{i\vk\cdot(\vu-\vu')}{|\vk \cdot (\vu - \vu')|^2 + \nu^2}\right],\\
	& & -\int_{t_0}^t d\tau\int d\vr'\int d\vr''\vec{F}(\vr-\vr') \vec{F}(\vr' - \vu'(t-\tau) - \vr'') \cdot \grad' \wz' \nonumber\\
	& & 	\times \left\langle \dwz(\vr'',\tau) \delta\wzpm \right\rangle\nonumber\\
	& = & \pm\Omega\int d\vr'\vec{F}(\vr - \vr')\int \frac{d\vk}{(2\pi)^2}\exp(i\vk\cdot(\vr - \vr')) \frac{\hat{\vec{z}}\times i\vk}{|\vk|^2}\cdot \wzpm \grad' \wz' \nonumber\\
	& & \times\left[\pi\delta(\vk \cdot (\vu - \vu')) - \frac{i\vk\cdot(\vu-\vu')}{|\vk \cdot (\vu - \vu')|^2 + \nu^2}\right].
\end{eqnarray}
The whole results are given by
\begin{eqnarray}
	\bGamma_{\pm}
	& = & -\Omega \int d\vr'\int \frac{d\vk}{(2\pi)^2} \int \frac{d\vk'}{(2\pi)^2}
		\exp(i(\vk+\vk')\cdot(\vr - \vr'))\nonumber\\
	& & \times \left[\pi\delta(\vk \cdot (\vu - \vu')) - \frac{i\vk\cdot(\vu-\vu')}{|\vk \cdot (\vu - \vu')|^2 + \nu^2}\right]\nonumber\\
	& & \times \frac{\hat{\vec{z}} \times i\vk'}{|\vk'|^2} \frac{\hat{\vec{z}} \times i\vk}{|\vk|^2} \cdot 
		\left[ (\wzp' - \wzm')\grad \wzpm \mp \wzpm \grad'\wz'\right] \label{eqn:gamma}
\end{eqnarray}
where we have used the relation
\begin{eqnarray}
	& & \vec{F}(\vr - \vq' - (\vu - \vu(\vq'))(t-\tau)) \nonumber\\
	& = & \frac{1}{(2\pi)^2} \int d\vk\frac{\hat{\vec{z}}\times i\vk}{|\vk|^2}\exp(i\vk\cdot(\vr - \vq' - (\vu - \vu(\vq'))(t-\tau))).\label{eqn:fourier}
\end{eqnarray}
It should be noted that the obtained diffusion fluxes (\ref{eqn:gamma}) can be divided into two parts,
namely the diffusion tensor ${\sf D}(\vr,t)$ proportional to $\grad \wzpm$
and the drift velocity $\vec{V}(\vr,t)$ proportional to $\wzpm$.
\begin{eqnarray}
	\bGamma_{\pm} & = & -{\sf D} \cdot \grad \wzpm \pm \vec{V}\wzpm, \label{eqn:diffusion_flux_pm}\\
	{\sf D} & = & \Omega \int d\vr'\int \frac{d\vk}{(2\pi)^2} \int \frac{d\vk'}{(2\pi)^2}
		\exp(i(\vk+\vk')\cdot(\vr - \vr'))\nonumber\\
	& & \times \left[\pi\delta(\vk \cdot (\vu - \vu')) - \frac{i\vk\cdot(\vu-\vu')}{|\vk \cdot (\vu - \vu')|^2 + \nu^2}\right]\nonumber\\
	& & \times \frac{\hat{\vec{z}} \times i\vk'}{|\vk'|^2} \frac{\hat{\vec{z}} \times i\vk}{|\vk|^2} (\wzp' - \wzm') \label{36}\\
	\vec{V} & = & \Omega \int d\vr'\int \frac{d\vk}{(2\pi)^2} \int \frac{d\vk'}{(2\pi)^2}
		\exp(i(\vk+\vk')\cdot(\vr - \vr'))\nonumber\\
	& & \times \left[\pi\delta(\vk \cdot (\vu - \vu')) - \frac{i\vk\cdot(\vu-\vu')}{|\vk \cdot (\vu - \vu')|^2 + \nu^2}\right]\nonumber\\
	& & \times \frac{\hat{\vec{z}} \times i\vk'}{|\vk'|^2} \frac{\hat{\vec{z}} \times i\vk}{|\vk|^2} \cdot \grad'\wz' \label{37}
\end{eqnarray}
Equations (\ref{36}) and (\ref{37}) include the oscillatory term $\exp(i(\vk+\vk')\cdot(\vr - \vr'))$.
To reveal the characteristics of the obtained collision term, 
we need to calculate the space average of the diffusion fluxes
to drop the high-frequency component.
Space average is calculated over the small rectangular area $\Lambda(\vr)$ with sides both $2L$
located at $\vec{r}$.
The space average of the diffusion fluxes $\bGamma_{\pm}$ is defined by
\begin{equation}
	\langle \bGamma_{\pm}\rangle_s \equiv \bGamma_{s\pm}(\vr) = \frac{1}{|\Lambda(\vr)|}\int_{\Lambda(\vr)}d\vr'' \bGamma_{\pm}(\vr'').
\end{equation}
We assume that the macroscopic variables such as $\vu$ and $\wz$ may be
constant inside $\Lambda(\vr)$ and only the term $\exp(i(\vk+\vk')\cdot(\vr - \vr'))$
should be space-averaged.

Finally, we obtain the following formulae for the diffusion and the drift terms.
\begin{eqnarray}
	\bGamma_{s\pm}(\vr) & \equiv & -{\sf D}_s \cdot \grad \wzpm \pm \vec{V}_s \wzpm,\label{eqn:final-gamma-app}\\
	{\sf D}_s & = & K \int d\vr' \frac{(\vu - \vu')(\vu-\vu')(\wzp' - \wzm')}{|\vu - \vu'|^3}, \label{eqn:final-D-app}\\
	\vec{V}_s & = & K \int d\vr' \frac{(\vu - \vu')(\vu-\vu')\cdot \grad'\omega'}{|\vu - \vu'|^3}, \label{eqn:final-V-app} \\
	K & = & \frac{\Omega}{(2\pi)^3} \left( \frac{\pi}{L}\right)^2\frac{1}{k_{\rm min}},
\end{eqnarray}
where the parameter $k_{\rm min}$ is introduced to regularize a singularity.
It is determined by the largest wave length that does not exceed the system
size, namely $k_{\rm min} = 2\pi / R$ where $R$ is a characteristic system
size determined by an initial distribution of the vortices.
Note that in (\ref{eqn:final-D-app}) and (\ref{eqn:final-V-app}), two unknown parameters 
$L$ and $k_{\rm min}$ remain.
\section*{References}

\end{document}